\pdfoutput=1
\documentclass[12pt]{article}
\usepackage{amsmath,amssymb,amsfonts}
\usepackage{tikz}
\begin{document}
\newtheorem{theorem}{Theorem}[section]
\newtheorem{corollary}[theorem]{Corollary}
\newtheorem{lemma}[theorem]{Lemma}
\newtheorem{remark}[theorem]{Remark}
\newtheorem{example}[theorem]{Example}
\newtheorem{proposition}[theorem]{Proposition}
\newtheorem{definition}[theorem]{Definition}

\def\emptyset{\varnothing}
\def\setminus{\smallsetminus}
\def\id{{\mathrm{id}}}
\def\C{{\mathbb{C}}}
\def\N{{\mathbb{N}}}
\def\Q{{\mathbb{Q}}}
\def\R{{\mathbb{R}}}
\def\Z{{\mathbb{Z}}}
\def\tr{{\mathrm{tr}}}
\def\a{{\alpha}}
\def\be{{\beta}}
\def\de{{\delta}}
\def\e{{\varepsilon}}
\def\si{{\sigma}}
\def\la{{\lambda}}
\def\th{{\theta}}
\def\lan{{\langle}}
\def\ran{{\rangle}}
\def\isom{{\cong}}
\newcommand{\Hom}{\mathop{\mathrm{Hom}}\nolimits}

\title{Two-dimensional topological order and operator algebras}

\author{
{\sc Yasuyuki Kawahigashi}\\
{\small Graduate School of Mathematical Sciences}\\
{\small The University of Tokyo, Komaba, Tokyo, 153-8914, Japan}\\
{\small e-mail: {\tt yasuyuki@ms.u-tokyo.ac.jp}}
\\[0,40cm]
{\small Kavli IPMU (WPI), the University of Tokyo}\\
{\small 5--1--5 Kashiwanoha, Kashiwa, 277-8583, Japan}
\\[0,40cm]
{\small Trans-scale Quantum Science Institute}\\
{\small The University of Tokyo, Bunkyo-ku, Tokyo 113-0033, Japan}
\\[0,05cm]
{\small and}
\\[0,05cm]
{\small iTHEMS Research Group, RIKEN}\\
{\small 2-1 Hirosawa, Wako, Saitama 351-0198,Japan}}
\maketitle{}

\begin{abstract}
We review recent interactions between mathematical theory of
two-dimensional topological order and operator algebras,
particularly the Jones theory of subfactors. 
The role of representation theory in terms of tensor categories
is emphasized.  Connections to 2-dimensional conformal field
theory are also presented.  In particular, we discuss anyon
condensation, gapped domain walls and matrix product operators
in terms of operator algebras.
\end{abstract}

\section{Introduction to operator algebras}

In quantum mechanics, an observable is represented by
a self-adjoint operator on a complex Hilbert space of states.  It would be
convenient if we are allowed to make algebraic operations such
as addition and multiplication on them,
so we consider a set of operators on a complex Hilbert space
which is closed under addition and multiplication.
In order to add and multiply them freely, we restrict our
attention to {\em bounded} linear operators, though many
observables appearing quantum mechanics are unbounded.
One way to get a bounded linear operator from an unbounded
one is exponentiation.  We exponentiate an unbounded 
self-adjoint operator $A$
to get a (bounded) unitary operator $\exp(iA)$.
Since self-adjointness of an observable is important, we also require
the set of operators is closed in the $*$-operation.
(In mathematics,
we write $A^*$ for the adjoint of $A$
rather than $A^\dagger$ which is more common
in physics literature.)
We further require that it is closed in a certain
topology, and then we call it an {\em operator
algebra}.

We have two natural choices for this topology.
One is the {\em norm topology} which arises from 
uniform convergence on the unit ball of the 
Hilbert space and the other is the {\em strong
operator topology} which arises from 
pointwise convergence on the Hilbert space.
(The strong operator topology is {\em weaker} than
the norm topology.)
When an operator algebra is closed in the norm
topology, it is called a {\em $C^*$-algebra}.
When an operator algebra is closed in the strong operator
topology, it is called a {\em von Neumann algebra}.
The latter was introduced by von Neumann in connection to
quantum mechanics and representation theory as the name
shows.  Both naturally appear in mathematical physics.
Being closed in a weaker topology is a stronger
condition, so a von Neumann algebra is automatically a 
$C^*$-algebra by definition, but it is usually not
very convenient to regard a von Neumann algebra as
a $C^*$-algebra since a typical von Neumann algebra is 
often quite different from a typical $C^*$-algebra.
For example, a typical $C^*$-algebra appearing
in applications is often separable (as a Banach
space), but a von Neumann algebra is never 
separable unless it is finite dimensional.

Finite dimensional $C^*$-algebras and
finite dimensional von Neumann algebras are
the same, and they are finite direct sums
of $M_n(\C)$, $n\times n$-matrix algebras with
complex entries. From this
viewpoint, one can say that an infinite dimensional
operator algebra is an infinite dimensional
analogue of a matrix algebra.

A commutative $C^*$-algebra is isomorphic
to $C(X)$ for some compact Hausdorff space $X$.
A commutative von Neumann algebra is
isomorphic to $L^\infty(Y,\mu)$ for some measure
space $(Y,\mu)$.  From this viewpoint, one can say
that a general operator algebra is a noncommutative
version of a function algebra.  We sometimes say
for this that a general operator algebra is
a function algebra on a
{\em noncommutative space}.  
Noncommutative geometry of Connes\cite{C} is a
far advancement of this idea.

A simple von Neumann algebra, in the sense that it
has only trivial closed two-sided ideals in the
strong operator topology, is called a {\em factor}.
A general von Neumann algebra is decomposed into a
direct integral of factors, a generalized version of a direct sum,
just like a representation decomposes into a
direct integral irreducible representations.  A
von Neumann algebra is a factor if and only if it does not
decompose into a direct sum of two nonzero von Neumann algebras.
A von Neumann algebra is a factor if and only if its
center is the complex number field $\C$.

The von Neumann algebra $B(H)$, the set of all bounded linear
operators on a Hilbert space $H$, is a factor, but this is
not very exciting.  A much more interesting factor is
constructed as follows.
A matrix $x\in M_n(\C)$ is mapped to $\left(\begin{array}{cc}
x & 0 \\
0 & x 
\end{array}\right)\in M_{2n}(\C)$.  With this embedding,
we have an increasing sequence of matrix algebras,
\[
M_2(\C)\subset M_4(\C)\subset M_8(\C)\subset\cdots.
\]
We take an increasing union $\bigcup_{n=1}^\infty M_{2^n}(\C)$.
This has a natural representation on a Hilbert space and we take
the closure of the image in the strong operator topology.
This gives a factor and is called the {\em hyperfinite II$_1$ factor}.

When one factor is $N$ contained in another factor $M$,
we say it is a {\em subfactor}.
Its study is somehow analogous to an algebraic study of
a subgroup $H\subset G$ and a subfield $K\subset L$.
We have a notion of the index $[G:H]$ for a
subgroup $H\subset G$ and the degree of extension
$[L:K]$ for a field extension $L\supset K$.
We have a similar notion of the
{\em Jones index} $[M:N]$ for a subfactor $N\subset M$.
One important new feature is that it is a positive real
number larger than or equal to 1, or infinity, and
can easily take a non-integer value.
We are interested in algebraic studies of
subfactors with finite Jones index.  Subfactors
of the hyperfinite II$_1$ factor have  very rich structures.

Theory of subfactors produced a new topological invariant,
the {\em Jones polynomial}\cite{J2} for {\em knots}, 
which is expected to be useful for topological quantum 
computations\cite{W}.

\section{Topological phases of matter and tensor categories}

We consider a certain 2-dimensional status of matters and
a typical example is thin liquid on a large plane.
A point on the plane can have a special status by
{\em excitation}.  An excited point behaves like
a particle ({\em quasi-particle}) and is called an {\em anyon}.  

Suppose we have finitely many anyons and study exchanges
among them.   The natural group for exchanges is
the {\em braid group} and we have braid group
statistics.  (If the original dimension where quasi-particles
live is 3, then the natural group representing
such an exchange is the permutation group,
and we have a {\em boson} or a {\em fermion}.)
An anyon is a more general
version of a boson and a fermion and we have this name 
because a phase arising from an exchange of quasi-particle
can take {\em any} value.
A {\em modular tensor category} gives a mathematical
description of such a system of finitely many anyons.
Each irreducible object of a modular tensor category
corresponds to an anyon.
Two anyons are fused and produce new anyons.  This is 
a fusion of anyons.
(See Bakalov-Kirillov\cite{BK}
for details of modular tensor categories.)  
A so-called non-abelian anyon is expected to produce a
topological quantum computer, but such an anyon has not been
observed experimentally yet.

Noncommutative geometry\cite{C} is a certain generalization
of Riemannian geometry using operator algebras.
An operator algebra with a certain extra structure
is regarded as a {\em noncommutative manifold}.
It has been used to study the {\em fractional quantum Hall
effect}.  This is also related to anyons.

A topological phase is also understood in terms of
homotopic classification of {\em gapped Hamiltonians} and we have a
version called {\em symmetry protected topological
phases}.  It has been recently studied extensively in the context of
$C^*$-algebras and index theorems by Ogata\cite{Og}.

We next explain the notion of a fusion category, which is a
more general version than a modular tensor category.
Recall that a finite group $G$ consists of the
following ingredients.
\begin{enumerate}
\item An associative multiplication
\item The identity element
\item The inverse elements
\end{enumerate}

For a finite group $G$, the set of its finite dimensional unitary
representations have the following structures.
\begin{enumerate}
\item Irreducible decomposition into finitely many ones
\item An associative tensor product
\item The identity representation
\item The contragredient representation
\end{enumerate}

Abstract axiomatization of such a set gives a notion of
a {\em fusion category}.  That is, each object behaves like
a representation, and we have only finitely many
irreducible objects.  We have direct sums and irreducible
decompositions, tensor products of objects, and the trivial
object and the dual object of each object, which behaves
like a contragredient representation.
We have morphisms between objects, like intertwiners
between representations.
For two representations $\pi$ and $\sigma$ of a group $G$,
the two tensor products $\pi\otimes\sigma$ and
$\sigma\otimes\pi$ are trivially unitarily equivalent,
but  such equivalence is {\em not} assumed 
for a general fusion category.  In this sense, the tensor
product operation in a fusion category is noncommutative
in general.  Fusion categories give
a special subclass of tensor categories.
(See Etingof-Nikshych-Ostrik\cite{ENO} for
a general theory of fusion categories. 
See Bischoff-Kawahigashi-Longo-Rehren\cite{BKLR} 
for tensor categories and operator algebras.)
Each object of a fusion category has a dimension, which is
a positive real number larger than or equal to 1.  A dimension
is additive with respect to a direct sum and multiplicative
with respect to a tensor product.

We have an important class of fusion categories for which
the above commutativity of tensor products holds in some
mathematically nice way.  Such commutativity is called
{\em braiding} because it is similar to reversing a
crossing of two wires.  (A wire labels an object in a 
fusion category.)
A braiding naturally comes in a pair --- overcrossing and
undercrossing.  It is more interesting if these two are
really different.  If this is the case in some appropriate
sense, this gives a notion of a {\em modular tensor category},
as mentioned above for anyons.
So objects in a modular tensor category behave more like
representations of a group, and they also behave like
particles.

The {\em Kitaev toric code}\cite{Ki}
gives one example of a modular
tensor category.  In this example, we have four 
irreducible objects representing four anyons.
Each irreducible object has dimension 1 in
this example, and the tensor product
rules (also called {\em fusion rules}) are given by the group
multiplication of
${\mathbb Z}/2{\mathbb Z}\times{\mathbb Z}/2{\mathbb Z}$.
(This group structure can give a trivial braiding, but here
we have a different, nontrivial braiding structure.)

Another example is the {\em Fibonacci category} and it
has two anyons, the trivial one labeled as $1$ and
another one labeled
as $\tau$.  The fusion rules are given by
$\tau^2=1\oplus\tau$.  This is expected to be
related to fractional quantum Hall liquids.
This is also related to the Jones
polynomial at the deformation parameter
$q=\exp(2\pi i/5)$.  The name Fibonacci comes from
the fact that certain intertwiner spaces have  dimensions
given by the Fibonacci numbers.

Suppose we have a modular tensor category having
$n$ irreducible objects.  The braiding produces an
$n$-dimensional unitary representation $\pi$ of
$SL(2,{\mathbb Z})$, the {\em modular} group. 
(See Bakalov-Kirillov\cite{BK}.)
We set $S=\displaystyle
\pi\left(
\begin{array}{cc}
0 & 1 \\
-1 & 0 
\end{array}
\right)$ and $T=\displaystyle
\pi\left(
\begin{array}{cc}
1 & 1 \\
0 & 1
\end{array}
\right)$.

A matrix $Z$ is called a {\em modular invariant}
if it satisfies the following, where the index $0$
means the trivial object, often called the {\em vacuum},
which behaves like a trivial representation.

\begin{enumerate}
\item $Z_{\lambda\mu}\in\{0,1,2,\dots\}$.
\item $Z_{00}=1$.
\item $ZS=SZ$, $ZT=TZ$.
\end{enumerate}

When a modular tensor category is given, the number
of modular invariants is always finite, and they
are sometimes explicitly classified.
See Cappelli-Itzykson-Zuber\cite{CIZ} for such a classification.
A typical appearance of a modular invariant is
2-dimensional conformal field theory.
Use of modular invariants in classification in
operator algebraic conformal field theory has been given
in Kawahigashi-Longo\cite{KL1}\cite{KL2}.

\section{Algebraic quantum field theory}

We now recall a mathematical (axiomatic) framework
of quantum field theory.  Our basic ingredients are
as follows (as in the {\em Wightman axioms}).

\begin{enumerate}
\item The spacetime
\item The spacetime symmetry group
\item Quantum fields (operator-valued distributions) on the spacetime
\end{enumerate}

This setting is well-studied and has a long history, but is
difficult to handle from a mathematical/technical
viewpoint, and we consider a different approach based on
operator algebras instead here.  We now consider a family of
von Neumann algebras parameterized by spacetime regions.
Each von Neumann algebra is generated by observables on
a spacetime region.  Though observables can be easily
unbounded operators, we consider von Neumann algebras of
bounded linear operators as before.  Our idea is that
relative positions  of this family of von Neumann algebras
encode physical information of a quantum field theory.
This approach is called {\em algebraic quantum field theory} 
as in Haag\cite{Ha}.

A 2-dimensional conformal field theory is a quantum field theory
with {\em conformal symmetry} on the 2-dimensional
Minkowski space.  It decomposes into an extension of a tensor
product of two quantum field theories living on the compactification
of the light
rays $\{x=\pm t\}$ and each is called a chiral conformal field theory 
(a {\em chiral half}).  It is a quantum field theory on the spacetime
$S^1$ with ${\mathrm{Diff}}(S^1)$-symmetry.  (Now the space and
the time are mixed into one dimension).

An operator algebraic formulation of a chiral conformal
field theory is called a (local)
{\em conformal net} in Kawahigashi-Longo\cite{KL1}.
This is based on operator algebras of
observables on $S^1$.  We have another mathematical formulation of
conformal field theory, based
on Fourier expansions of operator-valued distributions on
the circle $S^1$, which is called
a {\em vertex operator algebra}.  
(This name means an algebra of vertex operators and is not
directly related to operator algebras.)
See Kawahigashi\cite{K2}\cite{K5} as 
general references on conformal nets and
vertex operator algebras.
These two are supposed to be different mathematical
formulations of the same physical theory, so they should
be equivalent, at least under some nice assumptions.
We have shown that we can pass from a vertex operator
algebra to a local conformal net and come back under 
some mild assumptions as in Carpi-Kawahigashi-Longo-Weiner\cite{CKLW}.  
This relation of the two
has been much studied recently.

Under the standard set of axioms of a conformal net,
each von Neumann algebra
appearing in one is always isomorphic to
the unique Araki-Woods factor of type III$_1$.  So each
von Neumann algebra contains no information about a conformal
field theory, but as a family of von Neumann algebras, it
does contain information about conformal field theory.

Representation theory of a conformal net is called a theory of
{\em superselection sectors}.  Each representation (of the family
of von Neumann algebras on another Hilbert space) is
called a superselection sector as in  
Doplicher-Haag-Roberts\cite{DHR1}\cite{DHR2}.
They gave a tensor product structure there as a composition of
DHR endomorphisms and it has a
{\em braiding} in the setting of conformal field theory
as in Fredenhagen-Rehren-Schroer\cite{FRS1}\cite{FRS2}.  
Under some finiteness
assumption, called {\em complete rationality},
we get a {\em modular tensor category} 
of representations as in Kawahigashi-Longo-M\"uger\cite{KLM}.

Representation theory of a {\em vertex operator algebra} is
theory of {\em modules}.  We have a tensor product structure
and a braiding for modules.  Under some finiteness assumption,
we also get a {\em modular tensor category} 
of modules as in Huang\cite{H1}\cite{H2}.

Under nice identification of a local conformal net and a vertex
operator algebra as in
Carpi-Kawahigashi-Longo-Weiner\cite{CKLW}, 
Gui\cite{G} has identification of the
corresponding representation categories for many examples.

\section{Topological phases of matter and gapped domain walls}

Recall a 2-dimensional
topological order is described with a modular tensor
category.  Suppose we have two topological orders described with
two modular tensor categories ${\mathcal C}_1$ and
${\mathcal C}_2$.  The exterior tensor product
${\mathcal C}_1\boxtimes{\mathcal C}_2^{\mathrm{opp}}$,
where ``opp'' means reversing the braiding, gives
a new modular tensor category.  The number of new anyons
is the product of the two numbers of anyons.

We have a physical notion of a {\em gapped domain wall}
between the two topological orders and it is 
mathematically defined to be
an irreducible local {\em Lagrangian Frobenius algebra} with
the base space $\bigoplus Z_{\lambda\mu}
\lambda\boxtimes\bar\mu$ in 
${\mathcal C}_1\boxtimes{\mathcal C}_2^{\mathrm{opp}}$,
where $Z_{\lambda\mu}=0,1,2,\dots$ gives a
{\em generalized modular invariant}.  (See Kawahigashi\cite{K3}
for a mathematically precise formulation.)

Lan-Wang-Wen\cite{LWW} conjectured that if we have a generalized
modular invariant matrix
$Z$ with $Z_{00}=1$ for modular tensor categories
${\mathcal C}_1$ and ${\mathcal C}_2$, and the entries of
matrix $Z$ satisfy some inequalities about
{\em multiplicities},
then there would exist a corresponding gapped domain wall.

However, subfactor theory easily {\em disproves} this conjecture.
Actually, the charge conjugation matrix 
$(\delta_{\lambda\bar\mu})$ gives a counterexample
for some modular tensor category ${\mathcal C}_1={\mathcal C}_2$
arising from a finite group
by a recent work of Davydov\cite{D}.  We have also given a correct
form of the conjecture using Witt equivalence in
Kawahigashi\cite{K3}

In physics literature, we have a notion of 
{\em composition} of two gapped domain walls and its
irreducible decomposition.  We would like
to formulate this notion mathematically.

Suppose we have three topological orders described with
three modular tensor categories
${\mathcal C}_1$, ${\mathcal C}_2$ and ${\mathcal C}_3$,
respectively.
We further assume to have two irreducible local Lagrangian
Frobenius algebras with the base space
$\bigoplus Z^1_{\lambda\mu}\lambda\boxtimes \bar\mu$ 
in ${\mathcal C}_1\boxtimes {\mathcal C}_2^{\mathrm{opp}}$ and 
$\bigoplus Z^2_{\mu\nu}\mu\boxtimes \bar\nu$ 
in ${\mathcal C}_2\boxtimes {\mathcal C}_3^{\mathrm{opp}}$.

We would like to have a new Frobenius algebra with the 
base space $\bigoplus (\sum_\mu Z^1_{\lambda\mu}Z^2_{\mu\nu})
\lambda\boxtimes \bar\nu$.  That is, the matrix part is
just given by a {\em matrix multiplication}.

We can construct a Frobenius algebra with the base space
$\bigoplus (\sum_\mu Z^1_{\lambda\mu}Z^2_{\mu\nu})
\lambda\boxtimes \bar\nu$ by considering a tensor product
functor and taking an intermediate Frobenius algebra
but this is {\em reducible}
in general.  We have a notion of {\em irreducible
decomposition} of a Frobenius algebra corresponding to
irreducible decomposition of an operator algebra.

We show that after irreducible decomposition, each
Frobenius algebra is local and Lagrangian in Kawahigashi\cite{K4}.
Being Lagrangian is shown to be equivalent to 
modular invariance property in M\"uger\cite{M}.

On the matrix level, we thus have a decomposition
$\sum_\mu Z^1_{\lambda\mu}Z^2_{\mu\nu}=
\sum_i Z^{3,i}_{\lambda\nu}$.

\section{The $\alpha$-induction and anyon condensation}

For a modular tensor category ${\mathcal C}$
and a Frobenius algebra, we have a machinery of
{\em $\alpha$-induction}, similar to the induction procedure
in the classical representation theory.  If ${\mathcal C}$
corresponds to a rational conformal field theory and it
has an extension (like a conformal embedding), then the
vacuum representation of the larger conformal field theory
gives a commutative Frobenius algebra in ${\mathcal C}$.  
Our setting is
a generalization of this, and then from a representation
of the smaller conformal field theory, we get something like
a representation of the larger conformal field theory.
This process of $\alpha$-induction
depends on a choice of a positive or negative
braiding, and we use symbols
$\alpha^\pm_\lambda$ for the
induced ``representations'' arising from an object
$\lambda$ in the modular
tensor category $\mathcal C$.  See 
B\"ockenhauer-Evans-Kawahigashi\cite{BEK1}\cite{BEK2}
for an operator algebraic formulation.   The Frobenius
algebra corresponds to the {\em dual canonical endomorphism} in 
B\"ockenhauer-Evans-Kawahigashi\cite{BEK1}\cite{BEK2}.
The intersection
of the images of positive and negative $\alpha$-inductions
give a new modular tensor category.  In the case of an
extension of a conformal field theory like a
conformal embedding, this new modular tensor category is
the representation category of the larger conformal field 
theory.  That is, positive and negative $\alpha$-inductions
produce only something like a representation, but the intersection
of their images really give a genuine representation.
This procedure corresponds to
{\em condensation of anyons}, as in the
branching rule of Bais-Slingerland\cite{BS}, and
objects in the Frobenius algebra
corresponds to condensed anyons in Bais-Slingerland\cite{BS}.  
(If we have a commutative Frobenius algebra, then
the condensed anyons have to be bosons.)
They make a new vacuum in the new modular tensor category.

Many researchers
are interested in {\em commutative} Frobenius algebras, but
the results and methods in 
B\"ockenhauer-Evans-Kawahigashi\cite{BEK1}\cite{BEK2}
apply also to noncommutative Frobenius algebras.  
Commutativity of a Frobenius algebra is called {\em locality} in  
B\"ockenhauer-Evans-Kawahigashi\cite{BEK1}\cite{BEK2}, since this
corresponds to locality of an extended
chiral conformal field theory.  The results in
B\"ockenhauer-Evans-Kawahigashi\cite{BEK1}\cite{BEK2}
are stated in terms of operator
algebras, but they can be translated into a language of
modular tensor categories alone.

The $\alpha$-induction machinery produces a 
{\em modular invariant} 
$Z_{\lambda\mu}=\langle \alpha^+_\lambda,\alpha^-_\mu\rangle$,
where $\lambda,\mu$ are irreducible objects in $\mathcal C$
as in B\"ockenhauer-Evans-Kawahigashi\cite{BEK1} and 
an irreducible local Lagrangian Frobenius algebra in
${\mathcal C}\boxtimes{\mathcal C}^{\mathrm{opp}}$ as
in Rehren\cite{R}. 
The latter coincides with the one constructed in
Fr\"ohlich-Fuchs-Runkel-Schweigert\cite{FFRS} 
in a more categorical language, as shown in 
Bischoff-Kawahigashi-Longo\cite{BKL}.

A general fusion category has no braiding, but we have a 
general procedure called {\em the Drinfel$'$d center} 
as in Drinfel$'$d\cite{Dr} which
produces a new {\em modular} tensor category from a
given fusion category.

The $\alpha$-induction produces a new fusion category
for one choice of $+$ or $-$ braidings.  Its Drinfel$'$d 
center Drinfel$'$d\cite{Dr} is given by the tensor product of the original
modular tensor category and the extended one 
as in B\"ockenhauer-Evans-Kawahigashi\cite{BEK3}.
This mathematical result coincides with a result
called {\em boundary-bulk duality} in physical
literature like Kong\cite{Ko}.  See Table 1 of Kong\cite{Ko} 
for more interpretation
of $\alpha$-induction in terms of anyon condensation.

See Bischoff-Jones-Lu-Penneys\cite{BJLP} 
for another aspect of anyon condensation
and fusion categories.

\section{Tensor networks and matrix product operators}

\begin{figure}[h]
\begin{center}
\begin{tikzpicture}
\draw [thick] (2,1) circle (0.3);
\draw [thick] (1,1)--(1.7,1);
\draw (2,1)node{$v$};
\draw (1,1)node[above]{$j$};
\draw [thick] (4,1) circle (0.3);
\draw [thick] (3,1)--(3.7,1);
\draw [thick] (4.3,1)--(5,1);
\draw (4,1)node{$a$};
\draw (3,1)node[above]{$j$};
\draw (5,1)node[above]{$k$};
\end{tikzpicture}
\caption{A vector $v_j$ and a matrix $a_{jk}$}
\label{tensor}
\end{center}
\end{figure}
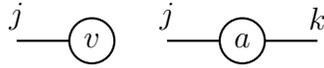

A vector $(v_j)$ has one index $j$.  This is pictorially represented
with one circle with one leg labeled with $j$ as in Fig. \ref{tensor}.
A matrix $(a_{jk})$ has two indices $j,k$.
This is pictorially represented with one circle with two legs 
labeled with $j, k$ as in Fig. \ref{tensor}.
Similarly, we can consider circles with 3 legs, 4 legs, and so on.
Such an object is called a {\em tensor}.

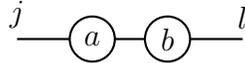
\begin{figure}[h]
\begin{center}
\begin{tikzpicture}
\draw [thick] (2,1) circle (0.3);
\draw [thick] (3,1) circle (0.3);
\draw [thick] (1,1)--(1.7,1);
\draw [thick] (2.3,1)--(2.7,1);
\draw [thick] (3.3,1)--(4,1);
\draw (2,1)node{$a$};
\draw (3,1)node{$b$};
\draw (1,1)node[above]{$j$};
\draw (4,1)node[above]{$l$};
\end{tikzpicture}
\caption{A matrix product $(ab)_{jl}$}
\label{matrix}
\end{center}
\end{figure}

The $(j,l)$ entry of the matrix product of $(a_{jk})$
and $(b_{kl})$ is given by $\sum_k a_{jk} b_{kl}$.  This is pictorially
represented with two circles corresponding to $(a_{jk})$
and $(b_{kl})$ where one leg of one and one leg of the other,
both labeled with $k$, are
concatenated as in Fig. \ref{matrix}.
Such concatenation is called a {\em contraction}.
We are interested in tensors with 3, 4 and 5 legs here.

\begin{figure}[h]
\begin{center}
\begin{tikzpicture}
\draw [thick] (2,2) circle (0.3);
\draw [thick] (3,2) circle (0.3);
\draw [thick] (4,2) circle (0.3);
\draw [thick] (1,2)--(1.7,2);
\draw [thick] (2.3,2)--(2.7,2);
\draw [thick,dashed] (3.3,2)--(3.7,2);
\draw [thick] (4.3,2)--(5,2);
\draw [thick] (1,1)--(3.3,1);
\draw [thick] (3.7,1)--(5,1);
\draw [thick,dashed] (3.3,1)--(3.7,1);
\draw [thick] (2,2.3)--(2,3);
\draw [thick] (3,2.3)--(3,3);
\draw [thick] (4,2.3)--(4,3);
\draw [thick] (1,2) arc (90:270:0.5);
\draw [thick] (5.5,1.5) arc (0:90:0.5);
\draw [thick] (5,1) arc (270:360:0.5);
\draw (2,2)node{$a$};
\draw (3,2)node{$a$};
\draw (4,2)node{$a$};
\draw (2,3)node[above]{$l_1$};
\draw (3,3)node[above]{$l_2$};
\draw (4,3)node[above]{$l_n$};
\draw (-0.5,2)node{$\displaystyle\sum_{l_1,l_2,\cdots,l_n}$};
\draw (7,2)node{$\left|l_1 l_2\cdots l_n\right\rangle$};
\end{tikzpicture}
\caption{A matrix product state}
\label{mps}
\end{center}
\end{figure}
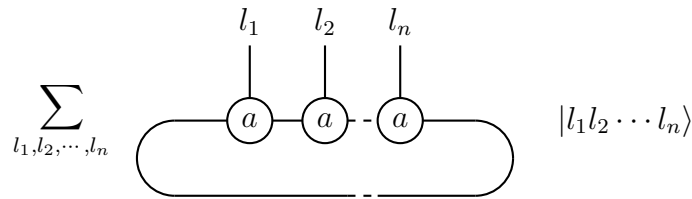

Consider a tensor with 3 legs and denote it by $a^l_{jk}$.
We consider a state represented as
\[
\sum_{l_1,l_2,\cdots,l_n}\mathrm{Tr}
(a^{l_1}a^{l_2}\cdots a^{l_n})\left|l_1 l_2\cdots l_n\right\rangle
\]
where $a^{l_m}$ is a matrix and a 3-tensor as in  
Fig. \ref{mps}.  Such a state was first
considered by Fannes-Nachtergaele-Werner\cite{FNW}
and is called a {\em matrix
product state (MPS)} today.

\begin{figure}[h]
\begin{center}
\begin{tikzpicture}
\draw [thick] (2,3) circle (0.3);
\draw [thick] (3,3) circle (0.3);
\draw [thick] (4,3) circle (0.3);
\draw [thick] (1,3)--(1.7,3);
\draw [thick] (2.3,3)--(2.7,3);
\draw [thick,dashed] (3.3,3)--(3.7,3);
\draw [thick] (4.3,3)--(5,3);
\draw [thick] (1,1)--(3.3,1);
\draw [thick] (3.7,1)--(5,1);
\draw [thick,dashed] (3.3,1)--(3.7,1);
\draw [thick] (2,3.3)--(2,4);
\draw [thick] (3,3.3)--(3,4);
\draw [thick] (4,3.3)--(4,4);
\draw [thick] (2,2)--(2,2.7);
\draw [thick] (3,2)--(3,2.7);
\draw [thick] (4,2)--(4,2.7);
\draw [thick] (1,3) arc (90:270:1);
\draw [thick] (6,2) arc (0:90:1);
\draw [thick] (5,1) arc (270:360:1);
\draw (-1.5,3)node{$\displaystyle\sum_{l_1,l_2,\dots,l_n,m_1,m_2,\dots,m_n}$};
\draw (2,3)node{$a$};
\draw (3,3)node{$a$};
\draw (4,3)node{$a$};
\draw (2,4)node[above]{$m_1$};
\draw (3,4)node[above]{$m_2$};
\draw (4,4)node[above]{$m_n$};
\draw (2,2)node[below]{$l_1$};
\draw (3,2)node[below]{$l_2$};
\draw (4,2)node[below]{$l_n$};
\draw (8,3)node{$\left|m_1 m_2\cdots m_n\right\rangle
\left\langle l_1 l_2 \cdots l_n\right|$};
\end{tikzpicture}
\caption{A matrix product operator}
\label{mpo}
\end{center}
\end{figure}
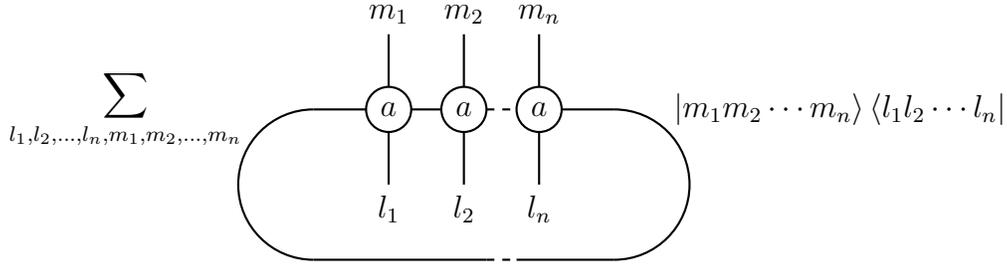

Next consider a tensor with 4 legs and denote it by
$b^{lm}_{jk}$.  We consider a matrix represented as
\[
\sum_{l_1,l_2,\dots,l_n,m_1,m_2,\dots,m_n}
\mathrm{Tr}(b^{l_1 m_1}\cdots b^{l_n m_n})
\left|l_1 l_2\cdots l_n\right\rangle
\left\langle m_1 m_2 \cdots m_n\right|
\]
as in Fig. \ref{mpo} and we call it a {\em matrix
product operator (MPO)}.

Bultinck-Mari\"ena-Williamson-\c Sahino\u glu-Haegeman-Verstraete\cite{BMWSHV} 
considered an algebra of matrix product operators
arising from a system of physically nice tensor networks of 4-tensors.
Such an algebra is called a {\em matrix product operator
algebra (MPOA)}.  They considered a category of
matrix product operators and presented a graphical method to
construct an interesting system of {\em anyons}.
They have shown that we get a {\em modular tensor category}
and discussed its physical consequence.

This paper has caught much attention of physicists.
They are aware of its similarity to an old work of Ocneanu\cite{EK1}.
We now show that their construction
is really  the same as Ocneanu's mathematically as in 
Kawahigashi\cite{K6}.

Let $N\subset M$ be a subfactor with finite Jones index
and finite depth. We set  $N=M_{-1}$, $M=M_0$ and apply the
Jones tower/tunnel construction  as in 
Definitions 9.24, 9.43 of Evans-Kawahigashi\cite{EK2} to get
\[
\cdots\subset M_{-2}\subset M_{-1}\subset M_0\subset M_1
\subset M_2\cdots,
\]
where the prime denotes the commutants.
We then consider a double sequence 
$A_{jk}=M'_{-k}\cap M_j$ of finite dimensional $C^*$-algebras.
They form {\em commuting squares} as in 
Section 9.6 of Evans-Kawahigashi\cite{EK2}.

Recall that a sequence
\[
\begin{array}{ccc}
A&\subset& B\\
\cap && \cap\\
C&\subset& D
\end{array}
\]
is called a commuting square if
the restriction of $E_B$ to $C$ is equal to $E_A$ where
$E_A$ and $E_B$ are {\em conditional expectations}
as in Theorem 5.25 of Evans-Kawahigashi\cite{EK2}.

Look at the Bratteli diagrams of a commuting square.  
(An inclusion of finite dimensional $C^*$-algebras gives
a Bratteli diagram as in page 67 of Evans-Kawahigashi\cite{EK2}.)
Choose one
edge from each diagram corresponding to each of the four inclusions.
We then get a complex number from the 4 edges.  This is a notion of a
{\em bi-unitary connection} as in Definition 11.3 of 
Evans-Kawahigashi\cite{EK2}
considered by Ocneanu and Haagerup.  (Also see Asaeda-Haagerup\cite{AH}.)
In the case as above,
we have an especially nice bi-unitary connection called a
{\em flat connection} as in Ocneanu\cite{O1}, 
Kawahigashi\cite{K1}, Section 11.4 of Evans-Kawahigashi\cite{EK2}.
(We also get this by fixing two labels of {\em quantum $6j$-symbols}.
See Chapter 12 of Evans-Kawahigashi\cite{EK2} for relations between quantum
$6j$-symbols and flat connections.)
If the original subfactor is hyperfinite and has finite Jones
index and finite depth, then this flat conenction
recovers the original subfactor entirely by Popa's theorem\cite{P1}.

We further have a finite system of flat connections corresponding
to the direct summands of $N'\cap M_{2n+1}$.
They produce a tensor of 4 legs.  
(We have the number 4 since a commuting
square has 4 inclusions.)  Though the normalization conventions
for a bi-unitary connection and a 4-tensor are slightly different,
the difference is only up to the Perron-Frobenius eigenvector
entries as in Fig. 11 of Kawahigashi\cite{K6}.
Such a tensor exactly fits
in the setting of 
Bultinck-Mari\"ena-Williamson-\c Sahino\u glu-Haegeman-Verstraete\cite{BMWSHV}.

Let all the four inclusion diagrams be the Dynkin diagram $A_n$.
Number the vertices as $1,2,\dots,n$ and set
$\varepsilon=\sqrt{-1}\exp\left(\pi\sqrt{-1}/{2(n+1)}\right)$.
We define a bi-unitary connection as in Fig. \ref{An},
where $\mu$ denotes the Perron-Frobenius weight, that is,
$\mu(j)=\sin(j\pi/(n+1))/\sin(\pi/(n+1))$.
(This also works for the Dynkin diagrams $D_n$, $E_6$, $E_7$
and $E_8$.  See Fig. 11.32 of Evans-Kawahigashi\cite{EK2}.)
This is actually a ``square root'' version of the one considered in 
Bultinck-Mari\"ena-Williamson-\c Sahino\u glu-Haegeman-Verstraete\cite{BMWSHV}.

\begin{figure}[h]
\begin{center}
\begin{tikzpicture}
\draw [thick, ->] (1,1)--(2,1);
\draw [thick, ->] (1,2)--(2,2);
\draw [thick, ->] (1,2)--(1,1);
\draw [thick, ->] (2,2)--(2,1);
\draw (1.5,1.5)node{$W$};
\draw (1,1)node[below left]{$l$};
\draw (1,2)node[above left]{$j$};
\draw (2,1)node[below right]{$m$};
\draw (2,2)node[above right]{$k$};
\draw (4.5,1.5)node{$\displaystyle=\delta_{kl}\e+
\sqrt{\frac{\mu(k)\mu(l)}{\mu(j)\mu(m)}}\delta_{jm}\bar\e$};
\end{tikzpicture}
\end{center}
\caption{A flat connection on the Dynkin diagram $A_n$}
\label{An}
\end{figure}

The above situations are very similar to 
interaction round the face (IRF) {\em solvable lattice models},
like the one due to Andrews-Baxter-Forrester\cite{ABF}.
For an IRF model, we fix a diagram, choose four edges from
it and make a square with them.  Then a complex value
called the {\em Boltzmann weight},
depending on a spectral parameter, is assigned to each such
square.  These values satisfy certain compatibility conditions
such as the {\em Yang-Baxter equation}.
The {\em crossing symmetry}
of an IRF model also corresponds to a flat connection.
When the diagram is the Dynkin diagram of type $A_n$,
our formula for the flat connection is essentially the
same as the restricted solid-on-solid (RSOS) model in Pasquier\cite{Pa}.

We have a construction of the {\em Drinfel$'$d center} as
in Drinfel$'$d\cite{Dr} which
gives a modular tensor category from a fusion category.  In
subfactor theory, the first such construction was
Ocneanu's {\em asymptotic inclusion} as in Evans-Kawahigashi\cite{EK1},
Section 12.6 of Evans-Kawahigashi\cite{EK2} which means
$N\vee (N'\cap M_\infty)\subset M_\infty$ constructed from
a hyperfinite II$_1$ subfactor $N\subset M$ with finite Jones
index and finite depth.  We later have
the Longo-Rehren subfactor construction\cite{LR} and Popa's
symmetric enveloping algebra construction\cite{P2}.

In connection to 3-dimensional topological quantum field
theory, Ocneanu further introduced a {\em tube algebra} 
as in Evans-Kawahigashi\cite{EK1}, Section 12.6 of 
Evans-Kawahigashi\cite{EK2}, Izumi\cite{I} which is
a finite dimensional $C^*$-algebra arising from a fusion
category.  Its each direct summand describes an
``anyon'' of the modular tensor category given
by the Drinfel$'$d center construction as in Drinfel$'$d\cite{Dr}.

Bultinck-Mari\"ena-Williamson-\c Sahino\u glu-Haegeman-Verstraete\cite{BMWSHV}
have a similar construction of the 
{\em anyon algebra} and it describes a system of anyons.

Suppose we start with a subfactor $N\subset M$
with finite Jones index and finite depth.
We then have a family of flat connections as above.  
(The number of flat connections
is equal to the dimension of the center of $N'\cap M_{2n+1}$ for
sufficiently large $n$.)  
We can next show that this family produces
a 4-tensor satisfying all the settings of
Bultinck-Mari\"ena-Williamson-\c Sahino\u glu-Haegeman-Verstraete\cite{BMWSHV}
simply by changing the normalization constants arising from 
the fourth root of the Perron-Frobenius
eigenvector entries as in Fig. 11 of Kawahigashi\cite{K6}.

\begin{theorem}
Ocneanu's tube algebra for the fusion category arising from a
subfactor and the anyon algebra of Bultinck et al.~arising 
from its flat connections are isomorphic.
In particular, the two fusion rules are identical and the 
Verlinde formula
also holds for the setting of anyon algebra.
\end{theorem}

We have seen that we can construct a tensor network satisfying
the requirements of 
Bultinck-Mari\"ena-Williamson-\c Sahino\u glu-Haegeman-Verstraete\cite{BMWSHV} 
from a subfactor.
It is known any  fusion category is realized from 
a subfactor.  Usually a ``representation''
of such a category is a bimodule
over II$_1$ factors or an endomorphism of a type III factor
as in Longo\cite{L1}\cite{L2}, but
we can also realize such a ``representation'' as a flat connection due
to the {\em open string bimodule} construction of Asaeda-Haagerup\cite{AH}.

\begin{theorem}
Suppose a tensor satisfies a setting of Bultinck et al.
If it realizes a fusion category, then this fusion category
is realized by a tensor of flat connections arising from
a subfactor as above.  Then the resulting
tube algebras and the modular tensor
categories are the same for the original and new systems.
\end{theorem}

In studies of topological phases, a {\em gapped Hamiltonian}
has been important.  It means a family of self-adjoint
matrices where the lowest eigenvalues are always separated
({\em ``gapped''}) from the next eigenvalues.  
Bultinck-Mari\"ena-Williamson-\c Sahino\u glu-Haegeman-Verstraete\cite{BMWSHV} 
studied such a
gapped Hamiltonian where the ground state is realized with a
{\em projected entangled-pair state (PEPS)}, which 
is a two-dimensional version of a matrix product state.
They have a certain expression of a matrix product operator,
which is called a projector matrix product operator (PMPO) since
it is also a projection, and the PEPS naturally lives in 
the range of several PMPOs.  We are thus interested in identification
of these PMPOs from an operator algebraic viewpoint.
(See Cirac-Perez-Garcia-Schuch-Verstraete\cite{CPGSV} 
for a general theory of PEPS.)

Our setting is as follows.  We have a 4-tensor corresponding to
a bi-unitary connection.  With this bi-unitary connection, we construct
a double sequence of finite-dimensional string algebras $A_{kl}$ with
$A_{00}=\C$, $A_{kl}\subset A_{k+1,l}$ and
$A_{kl}\subset A_{k,l+1}$ as in 
Section 11.3 of Evans-Kawahigashi\cite{EK2}
(with any choice of the starting vertex $*$)
and this produces hyperfinite II$_1$ factors 
$A_{\infty,l}$ and $A_{k,\infty}$ as the closures of
$\bigcup_k A_{kl}$ and $\bigcup_l A_{kl}$ in the strong
operator topology on appropriate Hilbert spaces.  
In particular, we have
two subfactors $A_{0,\infty}\subset A_{1,\infty}$
and $A_{\infty,0}\subset A_{1,\infty}$, both with finite
Jones index as in Theorem 11.9 of Evans-Kawahigashi\cite{EK2}.  
Suppose one of the two
subfactors has finite depth.  Then so does the other by Sato\cite{S1}.
In this case, we get a finite
family of bi-unitary connections $\{W_a\}$ as in Section 3
of Asaeda-Haagerup\cite{AH}, 
Section 2 of Kawahigashi\cite{K7}, corresponding to irreducible 
$A_{0,\infty}$-$A_{0,\infty}$ bimodules arising from the subfactor
$A_{0,\infty}\subset A_{1,\infty}$.
Each $W_a$ gives a corresponding 4-tensor
and they give a PMPO of length $k$ as in Section 3.1 of 
Bultinck-Mari\"ena-Williamson-\c Sahino\u glu-Haegeman-Verstraete\cite{BMWSHV}, Section 3 of Kawahigashi\cite{K7}.
We are interested in identifying the range of this projection.
(Note that the range of this projection is much smaller than
the entire Hilbert space.)  Then with subfactor technique based
on {\em flat fields of strings} of Ocneanu\cite{O2},
Theorem 11.15 of  Evans-Kawahigashi\cite{EK2},
and Definition 3 of Asaeda-Haagerup\cite{AH}
gives the following as in Theorem 3.3 of Kawahigashi\cite{K7}.

\begin{theorem}
The range of the $k$th PMPO in this setting
is naturally identified with the $k$th higher relative commutant
$A'_{\infty,0}\cap A_{\infty,k}$ for the subfactor
$A_{\infty,0}\subset A_{\infty,1}$ arising from the original 
bi-unitary connection.
\end{theorem}

The ranges of the $k$th PMPOs give an increasing sequence of
finite dimensional Hilbert spaces indexed by $k$ and they
are Hilbert spaces we would like to study in connection to
2-dimensional topological order.  The $k$th higher relative
commutants give an increasing sequence of finite dimensional
$C^*$-algebras with a positive definite inner product 
arising from a trace and they have been extensively studied
for almost 40 years in subfactor theory.
The above identification is expected to be useful to study
further relations between the two theories.

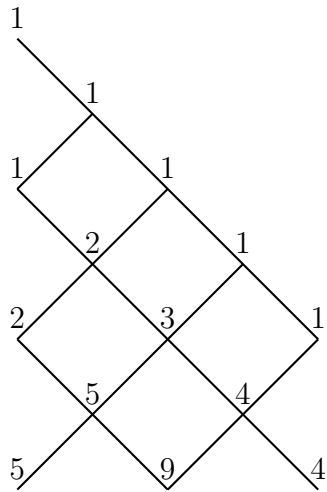
\begin{figure}[h]
\begin{center}
\begin{tikzpicture}
\draw [thick] (1,7)--(5,3);
\draw [thick] (1,5)--(5,1);
\draw [thick] (1,3)--(3,1);
\draw [thick] (1,5)--(2,6);
\draw [thick] (1,3)--(3,5);
\draw [thick] (1,1)--(4,4);
\draw [thick] (3,1)--(5,3);
\draw (1,7)node[above]{$1$};
\draw (1,5)node[above]{$1$};
\draw (1,3)node[above]{$2$};
\draw (1,1)node[above]{$5$};
\draw (2,6)node[above]{$1$};
\draw (2,4)node[above]{$2$};
\draw (2,2)node[above]{$5$};
\draw (3,5)node[above]{$1$};
\draw (3,3)node[above]{$3$};
\draw (3,1)node[above]{$9$};
\draw (4,4)node[above]{$1$};
\draw (4,2)node[above]{$4$};
\draw (5,3)node[above]{$1$};
\draw (5,1)node[above]{$4$};
\end{tikzpicture}
\caption{The Bratteli diagram of the higher relative commutants}
\label{bra}
\end{center}
\end{figure}

Take the Dynkin diagram $A_5$ as an example and consider the
bi-unitary connection as in Fig. \ref{An}.  Since the graph $A_5$ has four
edges, the dimension of the Hilbert space on which the $k$th PMPO
acts is $4^{2k}$.  Since this bi-unitary connection is flat, the Bratteli
diagram of the higher relative commutants is given as in 
Fig. \ref{bra}.  For example, the row for the
6th higher relative commutant says $5,9,4$, which means the
algebra is $M_5(\C)\oplus M_9(\C)\oplus M_4(\C)$.  (The top row
is counted as the 0th.)  The dimension
of this algebra is $5^2+9^2+4^2=122$.  An elementary computation
shows that the dimension of the $k$th higher relative
commutants is $(3^{k-1}+1)/2$ and this number is much smaller
than $16^k$.  If we take the Dynkin diagram $A_n$ for a large $n$,
the difference is even more drastic.

Note that the subfactor for the higher relative commutants here
is $A_{\infty,0}\subset A_{\infty,1}$  while we had the other
subfactor $A_{0,\infty}\subset A_{1,\infty}$ for the initial
finite depth assumption above.  The relation
between the two subfactor was studied in Sato\cite{S2}, and they
are characterized as mutually anti-Morita equivalent subfactors.
(In the above example of the Dynkin diagram $A_5$, the two
subfactors are the same.)

A recent general treatment of such tensors and tensor categories
is given in 
Lootens-Fuchs-Haegeman-Schweigert-Verstraete\cite{LFHSV}.  
Though the language is a little bit
different, this is also closely related to subfactor theory.

\section*{Acknowledgements}

This work was partially supported by 
JST CREST program JPMJCR18T6 and
Grants-in-Aid for Scientific Research 19H00640
and 19K21832.

\end{document}